\documentclass[italian,english]{article}
\usepackage[T1]{fontenc}
\usepackage[latin1]{inputenc}
\usepackage{color}
\usepackage{graphicx}
\usepackage{amssymb}

\makeatletter

 \newcommand{\lyxaddress}[1]{
   \par {\raggedright #1 
   \vspace{1.4em}
   \noindent\par}
 }

\usepackage{babel}
\makeatother
\begin{document}

\title{\textbf{Analysis of the transverse effect of Einstein's gravitational
waves }}

\author{\textbf{Christian Corda}}

\maketitle

\lyxaddress{\begin{center}INFN - Sezione di Pisa and Università di Pisa, Via
F. Buonarroti 2, I - 56127 PISA, Italy\end{center}}

\lyxaddress{\begin{center}\textit{E-mail address:} \textcolor{blue}{christian.corda@ego-gw.it} \end{center}}

\begin{abstract}
The investigation of the transverse effect of gravitational waves
(GWs) could constitute a further tool to discriminate among several
relativistic theories of gravity on the ground. After a review of
the TT gauge, the transverse effect of GWs arising by standard general
relativity (called Einstein's GWs in this paper) is reanalized with
a different choice of coordinates. In the chosen gauge test masses
have an apparent motion in the direction of propagation of the wave,
while in the transverse direction they appear at rest. Of course,
this is only a gauge artefact. In fact, from careful investigation
of this particular gauge, it is shown that the tidal forces associated
with GWs act along the directions orthogonal to the direction of propagation
of waves. In the analysis it is also shown, in a heuristic way, that
the transverse effect of Einstein's GWs arises from the propagation
of the waves at the speed of the light, thus only massless GWs are
purely transverse. But, because the physics of gravitational waves
has to be investigated by studing the tidal forces as appearing in
the geodesic deviation equation and directly in a laboratory environment
on Earth, an analysis of these tidal forces and of the transverse
effect in the frame of the local observer is also performed. After
this, for a further better understanding of the transverse effect,
an example of a wave, which arises from scalar tensor gravity, with
both transverse and genuinely longitudinal modes is given and discussed.
In the example the connection between the longitudinal component and
the velocity of the wave will be mathematical shown.

At the end of this paper the review of the TT gauge is completed,
recovering the gauge invariance between the presented gauge and the
TT one.
\end{abstract}

\lyxaddress{PACS numbers: 04.80.Nn, 04.30.Nk, 04.30.-w}

\section{Introduction}

The design and construction of a number of sensitive detectors for
GWs is underway today. There are some laser interferometers like the
VIRGO detector, being built in Cascina, near Pisa by a joint Italian-French
collaboration \cite{key-1,key-2}, the GEO 600 detector, being built
in Hannover, Germany by a joint Anglo-Germany collaboration \cite{key-3,key-4},
the two LIGO detectors, being built in the United States (one in Hanford,
Washington and the other in Livingston, Louisiana) by a joint Caltech-Mit
collaboration \cite{key-5,key-6}, and the TAMA 300 detector, being
built near Tokyo, Japan \cite{key-7,key-8}. There are many bar detectors
currently in operation too, and several interferometers and bars are
in a phase of planning and proposal stages.

The results of these detectors will have a fundamental impact on astrophysics
and gravitation physics. There will be lots of experimental data to
be analyzed, and theorists will be forced to interact with lots of
experiments and data analysts to extract the physics from the data
stream.

Detectors for GWs will also be important to confirm or ruling out
the physical consistency of general relativity or of any other theory
of gravitation \cite{key-9,key-10,key-11,key-12}. This is because,
in the context of extended theories of gravity, some differences from
general relativity and the others theories can be seen starting by
the linearized theory of gravity \cite{key-9,key-10,key-12,key-13,key-14,key-15}.
In fact, standard Einstein's GWs \cite{key-18} are in principle different
from GWs which arises from others theories of gravity. For example,
there are particular extended theories that admit the existence of
massive modes of gravitational radiation which can be both scalar
and tensorial \cite{key-9,key-12,key-13,key-14,key-15}. In this case,
the presence of the mass prevents the GW to propagate at the speed
of light generating a longitudinal component of the wave (i.e GWs
are no more transverse).

The response of interferometers to Einstein's GWs and their transverse
effect have been analyzed in lots of works in literature, starting
by the work of the Bondi's research group \cite{key-19} expecially
in the TT gauge \cite{key-20,key-21,key-22}. With the goal of considering
the investigation of the transverse effect of GWs a further tool to
discriminate among several relativistic theories of gravity on the
ground \cite{key-12,key-13,key-14,key-15,key-16,key-17}, in this
paper, after a review of the TT gauge, the transverse effect of Einstein's
GWs is reanalized with a different choice of coordinates. In the chosen
gauge test masses have an apparent motion in the direction of propagation
of the wave, while in the transverse direction they appear at rest.
Of course, this is only a gauge artefact. In fact, from careful investigation
of this particular gauge, it is shown that the tidal forces associated
with GWs act along the directions orthogonal to the direction of propagation
of waves. In the analysis it is also shown, in a heuristic way, that
the tranverse effect of Einstein's GWs arises from the propagation
of the waves at the speed of light, thus only massless GWs (and this
is the case of Einstein's ones) are purely transverse (i.e. the presence
of the mass precludes GWs to propagate at the speed of the light).
But, because the physics of gravitational waves has to be investigated
by studing the tidal forces as appearing in the geodesic deviation
equation \cite{key-25} and directly in a laboratory environment on
Earth\cite{key-12,key-13,key-14,key-15,key-20,key-21,key-22,key-24},
an analysis of these tidal forces and of the transverse effect in
the frame of the local observer is also performed, following the lines
of \cite{key-12,key-15,key-20}. After this, for a further better
understanding of the transverse effect, an example of a wave, which
arises from scalar tensor gravity, with both transverse and genuinely
longitudinal modes is given and discussed \cite{key-12,key-13,key-14}.
In the example the connection between the longitudinal component and
the velocity of the wave will be mathematical shown.

At the end of this paper the review of the TT gauge is completed recovering
the gauge invariance between the presented gauge and the TT one.

\section{A review of the TT gauge}

Working with $c=1$ and $\hbar=1$, in the context of General Relativity,
Einstein field equations can be written like \cite{key-21,key-23} 

\begin{equation}
G_{\mu\nu}\equiv R_{\mu\nu}-\frac{R}{2}\eta_{\mu\nu}=8\pi\tilde{G}T_{\mu\nu}^{(m)}.\label{eq: Einstein}\end{equation}

Because we want to study gravitational waves, the linearized theory
in vacuum ($T_{\mu\nu}^{(m)}=0$) has to be analyzed, with a little
perturbation of the background, which is assumed given by the Minkowskian
background \cite{key-21,key-23} .

Thus we put

\begin{equation}
g_{\mu\nu}=\eta_{\mu\nu}+h_{\mu\nu}\textrm{ with }\mid h_{\mu\nu}\mid\ll1\label{eq: linearizza}\end{equation}

To first order in $h_{\mu\nu}$, calling $\widetilde{R}_{\mu\nu\rho\sigma}$
, $\widetilde{R}_{\mu\nu}$ and $\widetilde{R}$ the linearized quantity
which correspond to $R_{\mu\nu\rho\sigma}$ , $R_{\mu\nu}$ and $R$,
the linearized field equations are obtained \cite{key-12,key-15,key-21,key-23}

\begin{equation}
\widetilde{R}_{\mu\nu}-\frac{\widetilde{R}}{2}\eta_{\mu\nu}=0.\label{eq: linearizzate1}\end{equation}

Let us put 

\begin{equation}
\bar{h}_{\mu\nu}\equiv h_{\mu\nu}-\frac{h}{2}\eta_{\mu\nu},\label{eq: h barra}\end{equation}

where the inverse transform is the same

\begin{equation}
h_{\mu\nu}=\bar{h}_{\mu\nu}-\frac{\bar{h}}{2}\eta_{\mu\nu}.\label{eq: h}\end{equation}

By putting eq. (\ref{eq: h}) in eqs. (\ref{eq: linearizzate1}) it
is

\begin{equation}
[]\bar{h}_{\mu\nu}-\partial_{\mu}(\partial^{\alpha}\bar{h}_{\alpha\nu})-\partial_{\nu}(\partial^{\alpha}\bar{h}_{\alpha\mu})+\eta_{\mu\nu}\partial^{\beta}(\partial^{\alpha}\bar{h}_{\alpha\beta}),\label{eq: onda}\end{equation}

where $[]$ is the D'Alembertian operator.

Now let us consider the gauge transform (Lorenz condition)

\begin{equation}
\bar{h}_{\mu\nu}\rightarrow\bar{h}'_{\mu\nu}=\bar{h}_{\mu\nu}-\partial_{(\mu}\epsilon_{\nu)}+\eta_{\mu\nu}\partial^{\alpha}\epsilon_{\alpha},\label{eq: gauge lorenzt}\end{equation}

with the condition $[]\epsilon_{\nu}=\partial^{\mu}\bar{h}_{\mu\nu}$
for the parameter $\epsilon^{\mu}$. It is

\begin{equation}
\partial^{\mu}\bar{h}'_{\mu\nu}=0,\label{eq: cond lorentz}\end{equation}

and, omitting the $'$, the field equations can be rewritten like

\begin{equation}
[]\bar{h}_{\mu\nu}=0.\label{eq: onda T}\end{equation}

For a lot of time in literature the condition (\ref{eq: gauge lorenzt})
was been called \textit{{}``Lorentz condition'',} because physicists
have considered its origin due to H. Lorentz (see for example refs.
\cite{key-21,key-23}), but its real origin is due to L. Lorenz, see
ref. \cite{key-26}. This particular was explicated in ref. \cite{key-17,key-27}
and was communicated to the author of this paper from \cite{key-28}.

Solutions of eqs. (\ref{eq: onda T}) are plan waves:

\begin{equation}
\bar{h}_{\mu\nu}=A_{\mu\nu}(\overrightarrow{k})\exp(ik^{\alpha}x_{\alpha})+c.c.\label{eq: sol T}\end{equation}

The solutions (\ref{eq: sol T}) take the conditions

\begin{equation}
\begin{array}{c}
k^{\alpha}k_{\alpha}=0\\
\\k^{\mu}A_{\mu\nu}=0,\end{array}\label{eq: vincoli}\end{equation}

which arises respectively from the linearized field equations and
from eq. (\ref{eq: cond lorentz}).

The first of eqs. (\ref{eq: vincoli}) shows that perturbations have
the speed of the light, the second represents the trasverse property
of the waves. Here this transverse effect appears from a purely mathematical
point of view, but in the next Section we show that this effect is
\textit{physical}.

Fixed the Lorenz gauge, another transformation with $[]\epsilon^{\mu}=0$
can be performed; let us take

\begin{equation}
\begin{array}{c}
[]\epsilon^{\mu}=0\\
\\\partial_{\mu}\epsilon^{\mu}=0,\end{array}\label{eq: gauge2}\end{equation}

which is permissed because $[]\bar{h}=0$. We obtain

\begin{equation}
\begin{array}{ccc}
\bar{h}=0 & \Rightarrow & \bar{h}_{\mu\nu}=h_{\mu\nu},\end{array}\label{eq: h ug h}\end{equation}

(traceless property) i.e. $h_{\mu\nu}$ is a transverse plane wave
too. The gauge transformation (\ref{eq: gauge2}) also saves the conditions

\begin{equation}
\begin{array}{c}
\partial^{\mu}\bar{h}_{\mu\nu}=0\\
\\\bar{h}=0.\end{array}\label{eq: vincoli 2}\end{equation}

Let us consider a wave incoming in the positive $z$ direction; then

\begin{equation}
k^{\mu}=(k,0,0k)\label{eq: k}\end{equation}

and the second of eqs. (\ref{eq: vincoli}) implies

\begin{equation}
\begin{array}{c}
A_{0\nu}=-A_{3\nu}\\
\\A_{\nu0}=-A_{\nu3}\\
\\A_{00}=-A_{30}+A_{33}.\end{array}\label{eq: A}\end{equation}

Now let us compute the freedom degrees of $A_{\mu\nu}$. We was started
with 10 components ($A_{\mu\nu}$ is a symmetric tensor); 3 components
have been lost for the trasverse property, more, the condition (\ref{eq: h ug h})
reduces the components to 6. One can take $A_{00}$, $A_{11}$, $A_{22}$,
$A_{21}$, $A_{31}$, $A_{32}$ like independent components; another
gauge freedom can be used to put to zero three more components (i.e.
one can only chose three of $\epsilon^{\mu}$, the fourth component
depends from the others by $\partial_{\mu}\epsilon^{\mu}=0$).

Then, taking

\begin{equation}
\begin{array}{c}
\epsilon_{\mu}=\tilde{\epsilon}_{\mu}(\overrightarrow{k})\exp(ik^{\alpha}x_{\alpha})+c.c.\\
\\k^{\mu}\tilde{\epsilon}_{\mu}=0,\end{array}\label{eq: ancora gauge}\end{equation}

the transform law for $A_{\mu\nu}$ is (see eqs. (\ref{eq: gauge lorenzt})
and (\ref{eq: sol T}) )

\begin{equation}
A_{\mu\nu}\rightarrow A'_{\mu\nu}=A_{\mu\nu}-2ik(_{\mu}\tilde{\epsilon}_{\nu}).\label{eq: trasf. tens.}\end{equation}

Thus, for the six components of interest it is:

\begin{equation}
\begin{array}{ccc}
A_{00} & \rightarrow & A_{00}+2ik\tilde{\epsilon}_{0}\\
A_{11} & \rightarrow & A_{11}\\
A_{22} & \rightarrow & A_{22}\\
A_{21} & \rightarrow & A_{21}\\
A_{31} & \rightarrow & A_{31}-ik\tilde{\epsilon}_{1}\\
A_{32} & \rightarrow & A_{32}-ik\tilde{\epsilon}_{2}.\end{array}\label{eq: sei tensori}\end{equation}

The physical components of $A_{\mu\nu}$ are the gauge-invariants
$A_{11}$, $A_{22}$ and $A_{21}$, thus $\tilde{\epsilon}_{\nu}$
can be chosen to put equal to zero the others. From the traceless
property it is also $A_{11}=-A_{22}.$

In this way the total perturbation of a gravitational wave propagating
in the $z+$ direction in this gauge is

\begin{equation}
h_{\mu\nu}(t-z)=A^{+}(t-z)e_{\mu\nu}^{(+)}+A^{\times}(t-z)e_{\mu\nu}^{(\times)},\label{eq: perturbazione totale}\end{equation}

that describes the two polarizations of gravitational waves which
arises from General Relativity in the TT gauge (ref. \cite{key-20,key-21,key-22}).
This gauge is historically called transverse-traceless (TT), because
in these particular coordinates the gravitational waves have a transverse
effect and are traceless.

\section{Einstein's GWs in a different gauge}

In the last section we have seen that, in the TT gauge, the perturbation
of a plane polarized Einstein's GW propagating in flat spacetime in
the $z+$ direction, with a wave front parallel to the $x-y$ plane,
is given , considering only the {}``$+$'' polarization, by 

\begin{equation}
h_{\mu\nu}(t-z)=h(t-z)e_{\mu\nu}^{(+)}\label{eq: perturbazione +}\end{equation}

(see also refs. \cite{key-20,key-21,key-22}), where $e_{\mu\nu}^{(+)}\equiv diag(0,1,-1,0)$,
the amplitude of the {}``$+$'' polarization is now labelled $h=h(t-z)$(i.e.
$A^{+}$ in previous Section ) and the line element is

\begin{equation}
ds^{2}=-dt^{2}+dz^{2}+[1+h(t-z)]dx^{2}+[1-h(t-z)]dy^{2}.\label{eq: metrica +}\end{equation}

A gauge transformation like eq. (\ref{eq: ancora gauge}) and the
corresponding transformation law (\ref{eq: trasf. tens.}) can also
be used to obtain $A_{00}=A_{22}$. Now the traceless property gives 

\begin{equation}
A_{11}=A_{33}=-A_{00}=-A_{22}\label{eq: AAAA}\end{equation}

and the line element is \begin{equation}
ds^{2}=[1+h(t-z)](-dt^{2}+dx^{2}+dz^{2})+[1-h(t-z)]dy^{2}.\label{eq: metrica + 3}\end{equation}

Equation (\ref{eq: metrica + 3}) can be also obtained directly from
the line element (\ref{eq: metrica +}) with the substitution

\begin{equation}
\begin{array}{ccc}
x & \rightarrow & x\\
\\y & \rightarrow & y\\
\\z & \rightarrow & z+\frac{1}{2}H(t-z)\\
\\t & \rightarrow & t-\frac{1}{2}H(t-z),\end{array}\label{eq: transf}\end{equation}

where

\begin{equation}
H(t-z)\equiv\int_{-\infty}^{t-z}h(v)dv.\label{eq: def H zero}\end{equation}

In literature it is well known that there exist infinitely many gauge
transformations that spoil the TT gauge \cite{key-21}. The particular
interest of the line element (\ref{eq: metrica + 3}) arises from
the fact that it is a TT gauge of an observer which is moving very
slowly in the $z$ direction as it will be shown in the next discussion.
In this case it will be also shown that this gauge artefact, that
arises in the relativity of motion, will generate a fallacious longitudinal
mode of the GW. More, two enlighten relations (eqs. (\ref{eq:  tempo di propagazione andata gauge Corda lungo z})
and (\ref{eq:  tempo di propagazione ritorno gauge Corda  lungo z}))
will show from an intuitive point of view that the tranverse effect
of Einstein's GWs arises from the propagation of the waves at the
speed of light.

Let us see in detail what happens in the gauge (\ref{eq: metrica + 3}).
The line element can be rewritten as

\begin{equation}
(\frac{dt}{d\tau})^{2}-(\frac{dx}{d\tau})^{2}-(\frac{dz}{d\tau})^{2}=\frac{1}{1+h}+\frac{1-h}{1+h}(\frac{dy}{d\tau})^{2}\label{eq: Ch2}\end{equation}

where $\tau$ is the proper time of the test masses.

To derive the geodesic equation of motion for test masses (i.e. the
beam-splitter and the mirrors of the interferometer) eq. (87,3a) of
\cite{key-11}, which is

\begin{equation}
\frac{du_{i}}{d\tau}-\frac{1}{2}\frac{\partial g_{kl}}{\partial x^{i}}u^{k}u^{l}=0,\label{eq: Landau geodesic}\end{equation}

can be used.

Thus, from the metric (\ref{eq: metrica + 3}) it is

\begin{equation}
\begin{array}{ccc}
\frac{d^{2}x}{d\tau^{2}} & = & 0\\
\\\frac{d^{2}y}{d\tau^{2}} & = & 0\\
\\\frac{d^{2}t}{d\tau^{2}} & = & \frac{1}{2(1+h)}\partial_{t}(1+h)[(\frac{dt}{d\tau})^{2}-(\frac{dx}{d\tau})^{2}-(\frac{dz}{d\tau})^{2}]-\frac{1}{2}\partial_{t}(1-h)(\frac{dy}{d\tau})^{2}\\
\\\frac{d^{2}z}{d\tau^{2}} & = & -\frac{1}{2(1+h)}\partial_{z}(1+h)[(\frac{dt}{d\tau})^{2}-(\frac{dx}{d\tau})^{2}-(\frac{dz}{d\tau})^{2}]+\frac{1}{2}\partial_{z}(1-h)(\frac{dy}{d\tau})^{2}.\end{array}\label{eq: geodetiche Corda*}\end{equation}

One can immediately integrate the first and the second of eqs. (\ref{eq: geodetiche Corda*})
obtaining

\begin{equation}
\frac{dx}{d\tau}=C_{1}=const.\label{eq: integrazione x}\end{equation}

\begin{equation}
\frac{dy}{d\tau}=C_{2}=const.\label{eq: integrazione x}\end{equation}

Assuming that test masses are at rest initially it is $C_{1}=C_{2}=0$.
Thus, even if the GW arrives at test masses, there is not motion of
test masses within the $x-y$ plane in this gauge. This can be directly
understood from eq. (\ref{eq: metrica + 3}) because the absence of
the $x$ and of the $y$ dependences in the metric implies that test
masses momentum in these directions (i.e. $C_{1}$ and $C_{2}$ respectively)
is conserved. This results, for example, from the fact that in this
case the $x$ and $y$ coordinates do not explicity enter in the Hamilton-Jacobi
equation for a test mass in a gravitational field (see ref. \cite{key-23}).

Now eq. (\ref{eq: Ch2}) reads

\begin{equation}
(\frac{dt}{d\tau})^{2}-(\frac{dz}{d\tau})^{2}=\frac{1}{1+h}.\label{eq: Ch3}\end{equation}

In this way, eqs. (\ref{eq: geodetiche Corda*}) begins

\begin{equation}
\begin{array}{ccc}
\frac{d^{2}x}{d\tau^{2}} & = & 0\\
\\\frac{d^{2}y}{d\tau^{2}} & = & 0\\
\\\frac{d^{2}t}{d\tau^{2}} & = & \frac{1}{2}\frac{\partial_{t}(1+h)}{(1+h)^{2}}\\
\\\frac{d^{2}z}{d\tau^{2}} & = & -\frac{1}{2}\frac{\partial_{z}(1+h)}{(1+h)^{2}}.\end{array}\label{eq: geodetiche Corda}\end{equation}

Now it will be shown that, in presence of a GW, a motion of test masses
in the $z$ direction, which is the direction of the propagating wave,
is present. An analysis of eqs. (\ref{eq: geodetiche Corda}) shows
that, to simplify equations, the retarded and advanced time coordinates
($v,w$) can be introduced:

\begin{equation}
\begin{array}{c}
v=t-z\\
\\w=t+z.\end{array}\label{eq: ret-adv}\end{equation}

From the third and the fourth of eqs. (\ref{eq: geodetiche Corda})
it is

\begin{equation}
\frac{d}{d\tau}\frac{dv}{d\tau}=\frac{\partial_{w}[1+h(v)]}{[1+h(v)]^{2}}=0.\label{eq: t-z t+z}\end{equation}

Thus one obtains

\begin{equation}
\frac{dv}{d\tau}=\alpha,\label{eq: t-z}\end{equation}

where $\alpha$ is an integration constant. From eqs. (\ref{eq: Ch3})
and (\ref{eq: t-z}), it is

\begin{equation}
\frac{dw}{d\tau}=\frac{\beta}{1+h}\label{eq: t+z}\end{equation}

where $\beta\equiv\frac{1}{\alpha}$, and

\begin{equation}
\tau=\beta v+\gamma,\label{eq: tau}\end{equation}

where the integration constant $\gamma$ correspondes simply to the
retarded time coordinate translation $v=t-z$. Thus, without loss
of generality, it can be put equal to zero. Now let us see what is
the meaning of the other integration constant $\beta$. The equation
for $z$ can be written from eqs. (\ref{eq: t-z}) and (\ref{eq: t+z}):

\begin{equation}
\frac{dz}{d\tau}=\frac{1}{2\beta}(\frac{\beta^{2}}{1+h}-1).\label{eq: z}\end{equation}

When it is $h=0$ (i.e. before the GW arrives at the test masses)
eq. (\ref{eq: z}) becomes\begin{equation}
\frac{dz}{d\tau}=\frac{1}{2\beta}(\beta^{2}-1).\label{eq: z ad h nullo}\end{equation}

But this is exactly the initial velocity of the test mass, thus we
have to choose $\beta=1$ because we suppose that test masses are
at rest initially. This also implies $\alpha=1$.

To find the motion of a test mass in the $z$ direction one can see
that from eq. (\ref{eq: tau}) it is $d\tau=dv$, while from eq. (\ref{eq: t+z})
it is $dw=\frac{d\tau}{1+h}$. Because it is $z=\frac{w-v}{2}$ we
obtain

\begin{equation}
dz=\frac{1}{2}(\frac{d\tau}{1+h}-dv),\label{eq: dz}\end{equation}

which can be integrated as

\begin{equation}
\begin{array}{c}
z=z_{0}+\frac{1}{2}\int(\frac{dv}{1+h}-dv)=\\
\\=z_{0}-\frac{1}{2}\int_{-\infty}^{t-z}\frac{h(v)}{1+h(v)}dv,\end{array}\label{eq: moto lungo z}\end{equation}

where $z_{0}$ is the initial position of the test mass. Now, the
displacement of the test mass in the $z$ direction can be written
as

\begin{equation}
\begin{array}{c}
\Delta z=z-z_{0}=-\frac{1}{2}\int_{-\infty}^{t-z_{0}-\Delta z}\frac{h(v)}{1+h(v)}dv\\
\\\simeq-\frac{1}{2}\int_{-\infty}^{t-z_{0}}\frac{h(v)}{1+h(v)}dv.\end{array}\label{eq: spostamento lungo z}\end{equation}

The results can also be rewritten in function of the time coordinate
$t$:

\begin{equation}
\begin{array}{ccc}
x(t) & = & x_{0}\\
\\y(t) & = & y_{0}\\
\\z(t) & = & z_{0}-\frac{1}{2}\int_{-\infty}^{t-z_{0}}\frac{h(v)}{1+h(v)}d(v)\\
\\\tau(t) & = & t-z(t).\end{array}\label{eq: moto gauge Corda}\end{equation}

Then, let us reasume what happens in our gauge: in the $x-y$ plane
an inertial test mass initially at rest remains at rest throughout
the entire passage of the GW, while in the $z$ direction an inertial
test mass initially at rest has a motion during the passage of the
GW. This motion has a very weak velocity, in fact from the third of
eqs. (\ref{eq: moto gauge Corda}) it is

\begin{equation}
\frac{dz}{dt}\simeq-\frac{h(t)}{2},\label{eq:velocity}\end{equation}

where $\frac{h(t)}{1+h(t)}\simeq h(t)$ has been used .

The following analysis will show that the fallacious effect of this
Section is only a gauge artefact. In fact it is well known from the
literature that GWs which arises from standard general relativity
are longitudinal waves \cite{key-2,key-9,key-11,key-18,key-19,key-20,key-21,key-22,key-23,key-24,key-25}.

\section{Analysis of the transverse effect in the chosen gauge}

The use of words {}`` at rest'' has to be clarified: it means that
the coordinates of test masses do not change in the presence of the
GW in the $x-y$ plane, but we will see that the proper distance between
the beam-splitter and the mirror of the interferometer changes even
though their coordinates remain the same. On the other hand, it will
also be shown that the proper distance between the beam-splitter and
the mirror of the interferometer does not change in the $z$ direction
even if their coordinates change in the gauge (\ref{eq: metrica + 3}). 

A good way to analyze variations in the proper distance (time) is
by means of {}``bouncing photons'' : a photon can be launched from
the beam-splitter to be bounced back by the mirror (see refs. \cite{key-12,key-15,key-20}
and figure 1). 

\begin{figure}
\includegraphics{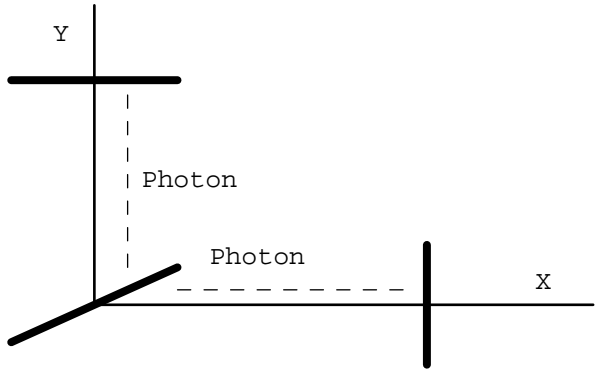}

\caption{photons can be launched from the beam-splitter to be bounced back
by the mirror}
\end{figure}

Let us start by considering the interval for a photon which propagates
in the $x$ axis. Assuming that test masses are located along the
$x$ axis and the $z$ axis of the coordinate system, the $y$ direction
can be neglected because the absence of the $y$ dependence in the
metric (\ref{eq: metrica + 3}) implies that photon momentum in this
direction is conserved \cite{key-23} and the interval can be rewritten
in the form

\begin{equation}
ds^{2}=[1+h(t-z)](-dt^{2}+dx^{2}+dz^{2}).\label{eq: metrica + 4}\end{equation}

Note that photon momentum in the $z$ direction is not conserved,
for the $z$ dependence in eq. (\ref{eq: metrica + 3}) \cite{key-20,key-23}.
Thus photons launched in the $x$ axis will deflect out of this axis.
But this effect can be neglected too, because the photon deflection
into the $z$ direction will be at most of order $h$ \cite{key-20}.
Then, to first order in $h$, the $dz^{2}$ term can be neglected.
Thus, from eq. (\ref{eq: metrica + 4}) it is

\begin{equation}
ds^{2}=(1+h)(-dt^{2})+(1+h)dx^{2}.\label{eq: metrica + 3 lungo x}\end{equation}

The condition for null geodesics ($ds^{2}=0$) for photons gives 

\begin{equation}
dt^{2}=dx^{2}.\label{eq: metrica puramente piu' di Corda lungo x 2}\end{equation}

One recalls that the rate $d\tau$ of the proper time is related to
the rate $dt$ of the time coordinate from (ref. \cite{key-23})

\begin{equation}
d\tau^{2}=g_{00}dt^{2}.\label{eq: relazione temporale}\end{equation}

From eq. (\ref{eq: metrica + 3 lungo x}) it is $g_{00}=(1+h)$. Then,
by using eq. (\ref{eq: metrica puramente piu' di Corda lungo x 2}),
one obtains

\begin{equation}
d\tau^{2}=(1+h)dx^{2},\label{eq: relazione spazial-temporale}\end{equation}

which gives 

\begin{equation}
d\tau=\pm(1+h)^{\frac{1}{2}}dx.\label{eq: relazione temporale 2}\end{equation}

From eqs. (\ref{eq: moto gauge Corda}) we see that the coordinates
of the beam-splitter $x_{b}=l$ and of the mirror $x_{m}=l+L_{0}$
do not change under the influence of the GW in our gauge, thus the
proper duration of the forward trip is

\begin{equation}
\tau_{1}(t)=\int_{l}^{L_{0}+l}[1+h(t)]^{\frac{1}{2}}dx.\label{eq: relazione temporale 2}\end{equation}

To first order in $h$ this integral is approximated by

\begin{equation}
\tau_{1}(t)=T_{0}+\frac{1}{2}\int_{l}^{L_{0}+l}h(t')dx\label{eq: durata volo andata approssimata in Corda x}\end{equation}

where

\begin{center}$t'=t-(l+L_{0}-x)$.\end{center}

In the last equation $t'$ is the retardation time (i.e. $t$ is the
time at which the photon arrives in the position $l+L_{0}$, so $l+L_{0}-x=t-t'$)
\cite{key-2,key-12,key-15,key-20}.

In the same way, the proper duration of the return trip is 

\begin{equation}
\tau_{2}(t)=T_{0}+\frac{1}{2}\int_{l+L_{0}}^{l}h(t')(-dx),\label{eq: durata volo ritorno approssimata in Corda x}\end{equation}

where now 

\begin{center}$t'=t-(x-l)$\end{center}

is the retardation time and

\begin{center}$T_{0}=L_{0}$ \end{center}

is the transit proper time of the photon in absence of the GW, which
also corresponds to the transit coordinate time of the photon in presence
of the GW (see eq. (\ref{eq: metrica puramente piu' di Corda lungo x 2})).

Thus the round-trip proper time will be the sum of $\tau_{2}(t)$
and $\tau_{1}(t-T_{0})$. Then, to first order in $h$, the proper
duration of the round-trip will be

\begin{equation}
\tau_{r.t.}(t)=\tau_{1}(t-T_{0})+\tau_{2}(t).\label{eq: durata round trip}\end{equation}

By using eqs. (\ref{eq: durata volo andata approssimata in Corda x})
and (\ref{eq: durata volo ritorno approssimata in Corda x}) one sees
immediately that deviations of this round-trip proper time (i.e. proper
distance) from its imperurbated value are given by

\begin{equation}
\delta\tau(t)=\frac{1}{2}\int_{l}^{L_{0}+l}[h(t-2T_{0}+x-l)+h(t-x+l)]dx.\label{eq: variazione temporale in gauge comovente}\end{equation}

The signal seen from the arm in the $x$ axis can be also defined
like

\begin{equation}
\frac{\delta\tau(t)}{T_{0}}\equiv\frac{1}{2T_{0}}\int_{l}^{L_{0}+l}[h(t-2T_{0}+x-l)+h(t-x+l)]dx.\label{eq: signal}\end{equation}

Now the analysis will be transled in the frequency domain by using
the Fourier transform of our field defined by

\begin{equation}
\tilde{h}(\omega)=\int_{-\infty}^{\infty}dt\textrm{ }h(t)\exp(i\omega t).\label{eq: trasformata di fourier}\end{equation}

By using definition (\ref{eq: trasformata di fourier}), from eq.
(\ref{eq: signal}) it is

\begin{equation}
\frac{\delta\tilde{\tau}(\omega)}{T_{0}}=\Upsilon(\omega)\tilde{h}(\omega),\label{eq: fourier in gage comoventi e Corda}\end{equation}

where $\Upsilon(\omega)$ is the response of the $x$ arm of our interferometer
to GWs:

\begin{equation}
\Upsilon(\omega)=\frac{\exp(2i\omega T_{0})-1}{2i\omega T_{0}},\label{eq: risposta in gages comovente e Corda}\end{equation}

which is computed in lots of works in literature, but here the computation
has been made in a different gauge.

Now let us see what happens in the $z$ coordinate (see figure 2).
\begin{figure}
\includegraphics{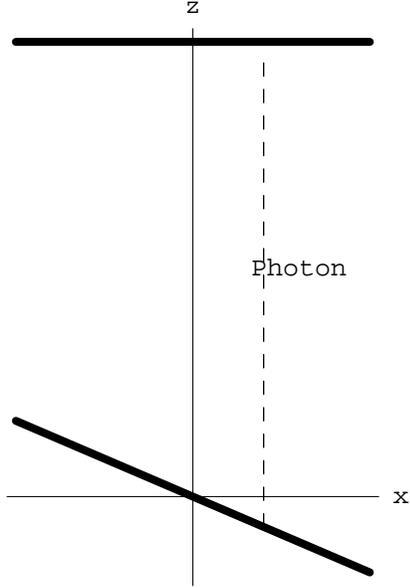}

\caption{the beam splitter and the mirror are located in the direction of
the incoming GW}
\end{figure}

The absences of the $x$ and $y$ dependence in the metric (\ref{eq: metrica + 3})
imply that photon momentum in these direction is conserved \cite{key-23}
and the interval can be rewritten in the form \begin{equation}
ds^{2}=[1+h(t-z)](-dt^{2}+dz^{2}).\label{eq: metrica z}\end{equation}

From the condition $ds^{2}=0$ for null geodesics it is

\begin{equation}
dz=\pm dt.\label{eq: metrica puramente piu di Corda lungo z 2}\end{equation}

But, from the last of eqs. (\ref{eq: moto gauge Corda}) the proper
time is

\begin{equation}
d\tau(t)=dt-dz,\label{eq: tempo proprio lungo z in Corda}\end{equation}

and, by combining eq. (\ref{eq: metrica puramente piu di Corda lungo z 2})
with eq. (\ref{eq: tempo proprio lungo z in Corda}), one gets

\begin{equation}
d\tau(t)=dt\mp dt.\label{eq: tempo proprio lungo z in Corda 2}\end{equation}

Thus, it is

\begin{equation}
\tau_{1}(t)=0\label{eq:  tempo di propagazione andata gauge Corda lungo z}\end{equation}

for the forward trip and

\begin{equation}
\tau_{2}(t)=\int_{0}^{T_{0}}2dt=2T_{0}\label{eq:  tempo di propagazione ritorno gauge Corda  lungo z}\end{equation}

for the return trip. Then

\begin{equation}
\tau(t)=\tau_{1}(t)+\tau_{2}(t)=2T_{0}.\label{eq: tempo proprio totale lungo z in Corda}\end{equation}

Thus it is $\delta\tau=\delta L_{0}=0$, i.e. there is not longitudinal
effect. This is a consequence of the fact that a GW propagates at
the speed of light. Eqs. (\ref{eq:  tempo di propagazione andata gauge Corda lungo z})
and (\ref{eq:  tempo di propagazione ritorno gauge Corda  lungo z})
show that, in the gauge (\ref{eq: metrica + 3}), in the forward trip
the photon travels at the same speed of the GW and its proper time
is equal to zero ((i.e. the photon and the GW are leaving at the same
velocity and in the same direction, giving the result (\ref{eq:  tempo di propagazione andata gauge Corda lungo z})),
while in the return trip the photon travels against \textbf{}the GW
and its proper time redoubles (i.e. the photon and the GW are leaving
at the same velocity, but in opposite direction, giving the result
(\ref{eq:  tempo di propagazione ritorno gauge Corda  lungo z})). 

Then it has been shown that the longitudinal effect of eqs. (\ref{eq: moto gauge Corda})
and (\ref{eq:velocity}) is fallacious and due to the relativity of
motion. The gauge (\ref{eq: metrica + 3}) is a TT gauge of an observer
which is moving very slowly in the $z$ direction.

\section{Tidal forces and geodesic deviation equations in the gauge of the
local observer}

Because the physics of GWs is directly performed in a laboratory environment
on Earth \cite{key-12,key-15,key-20,key-21,key-25}, the coordinate
system in which the space-time is locally flat is typically used and
the distance between any two points is given simply by the difference
in their coordinates in the sense of Newtonian physics. This gauge
is the proper reference gauge of a local observer, located for example
in the position of the beam splitter of an interferometer. In this
gauge GWs manifest themself by exerting tidal forces on the masses
\cite{key-25} (the mirror and the beam-splitter in the case of an
interferometer). A detailed analysis of the frame of the local observer
is given in ref. \cite{key-21}, sect. 13.6. Here the more important
features of this gauge are recalled:

the time coordinate $x_{0}$ is the proper time of the observer O;

spatial axes are centered in O;

in the special case of zero acceleration and zero rotation the spatial
coordinates $x_{j}$ are the proper distances along the axes and the
frame of the local observer reduces to a local Lorentz frame: in this
case the line element reads \cite{key-21}

\begin{equation}
ds^{2}=-(dx^{0})^{2}+\delta_{ij}dx^{i}dx^{j}+O(|x^{j}|^{2})dx^{\alpha}dx^{\beta};\label{eq: metrica local lorentz}\end{equation}

the effect of the GW on test masses is described by the equation

\begin{equation}
\ddot{x^{i}}=-\widetilde{R}_{0k0}^{i}x^{k},\label{eq: deviazione geodetiche}\end{equation}
which is the equation for geodesic deviation in this gauge. The problem
is thus reduced to calculate the linearized Riemann tensor $\widetilde{R}_{0k0}^{i}$
in the local gauge of the beam-splitter. But it is well known that
$\widetilde{R}_{0k0}^{i}$ is gauge invariant \cite{key-12,key-15,key-20,key-21},
thus we can compute it directly from eq. (\ref{eq: metrica + 3}).
From \cite{key-21} it is\begin{equation}
\widetilde{R}_{\mu\nu\rho\sigma}=\frac{1}{2}\{\partial_{\mu}\partial_{\beta}h_{\alpha\nu}+\partial_{\nu}\partial_{\alpha}h_{\mu\beta}-\partial_{\alpha}\partial_{\beta}h_{\mu\nu}-\partial_{\mu}\partial_{\nu}h_{\alpha\beta}\}.\label{eq: riemann lineare}\end{equation}
In this way one obtains, \begin{equation}
\begin{array}{c}
\widetilde{R}_{010}^{1}=-\frac{1}{2}\ddot{h}\\
\\\widetilde{R}_{010}^{2}=\frac{1}{2}\ddot{h}\\
\\\widetilde{R}_{030}^{3}=0.\end{array}\label{eq: componenti riemann}\end{equation}

Using eqs. (\ref{eq: deviazione geodetiche}), eqs: (\ref{eq: componenti riemann})
give \begin{equation}
\ddot{x}=\frac{1}{2}\ddot{h}x,\label{eq: accelerazione mareale lungo x}\end{equation}

\begin{equation}
\ddot{y}=-\frac{1}{2}\ddot{h}y\label{eq: accelerazione mareale lungo y}\end{equation}

and 

\begin{equation}
\ddot{z}=0,\label{eq: accelerazione mareale lungo z}\end{equation}

which show that the tidal forces act only in a direction perpendicular
to the propagating GW, while there is not longitudinal tidal force.

\section{An example of real longitudinal effect in a non - Einsteinian GW}

In this Section it is shown that Scalar-Tensor Gravity, which emerges
as {}``effective theory'' from several unification schemes of fundamental
interactions (see refs. \cite{key-10,key-12,key-13,key-14} for a
discussion), admits the existence of a scalar massive mode of gravitational
waves, where the mass is very small. The mechanism of production is
also analyzed, showing that this scalar massive mode has a real (no
gauge artefact) longitudinal mode \cite{key-12,key-14}. In this way
the correlation between the longitudinal mode and the fact that a
GW does not propagate at the speed of light will be made clear also
from a mathematical point of view.

Let us ask: what does it mean the term {}``small''? If one treats
scalars like classical waves, that act coherently with the interferometer
\cite{key-12,key-13,key-14}, it has to be $m\ll1/L$ , where $L=3$
kilometers in the case of Virgo and $L=4$ kilometers in the case
of Ligo. Thus it is approximately $m<10^{-9}eV$. However there is
a stronger limitation coming from the fact that the scalar wave should
have a frequency which falls in the frequency-range for earth based
gravitational antennas that is the interval $10Hz\leq f\leq10KHz$.
For a massive SGW this means:

\begin{equation}
2\pi f=\omega=\sqrt{m^{2}+p^{2}},\label{eq: frequenza-massa}\end{equation}

were $p$ is the momentum. Then it has to be \cite{key-12,key-13,key-14}

\begin{equation}
0eV\leq m\leq10^{-11}eV.\label{eq: range di massa}\end{equation}

For these light scalars we can still discuss their effect as a coherent
gravitational wave. 

In the general context of Scalar-Tensor Gravity \cite{key-10,key-12,key-13,key-14},
Einstein field equations are more general than eq. (\ref{eq: Einstein}): 

\begin{equation}
\begin{array}{c}
G_{\mu\nu}=-\frac{4\pi\tilde{G}}{\varphi}T_{\mu\nu}^{(m)}+\frac{\omega(\varphi)}{\varphi^{2}}(\varphi_{;\mu}\varphi_{;\nu}-\frac{1}{2}g_{\mu\nu}g^{\alpha\beta}\varphi_{;\alpha}\varphi_{;\beta})+\\
\\+\frac{1}{\varphi}(\varphi_{;\mu\nu}-g_{\mu\nu}[]\varphi)+\frac{1}{2\varphi}g_{\mu\nu}W(\varphi)\end{array}\label{eq: einstein-general}\end{equation}

with associed a Klein - Gordon equation for the scalar field

\begin{equation}
[]\varphi=\frac{1}{2\omega(\varphi)+3}(-4\pi\tilde{G}T^{(m)}+2W(\varphi)+\varphi W'(\varphi)+\frac{d\omega(\varphi)}{d\varphi}g^{\mu\nu}\varphi_{;\mu}\varphi_{;\nu}.\label{eq: KG}\end{equation}
In the above equations $T_{\mu\nu}^{(m)}$ is the ordinary stress-energy
tensor of the matter and $\tilde{G}$ is a dimensional, strictly positive,
constant \cite{key-10,key-12}. The Newton constant is replaced by
the effective coupling

\begin{equation}
G_{eff}=-\frac{1}{2\varphi},\label{eq: newton eff}\end{equation}

which is, in general, different from $G$. General Relativity is obtained
when the scalar field coupling becomes

\begin{equation}
\varphi=const=-\frac{1}{2}.\label{eq: varphi}\end{equation}

The case in which it is $\omega=const$ in eqs. (\ref{eq: einstein-general})
and (\ref{eq: KG}) is the string-dilaton gravity \cite{key-10,key-12,key-13,key-14}.
Now the linearized theory in vacuum ($T_{\mu\nu}^{(m)}=0$) has to
be studied in a little different perturbation of the background with
respect standard Einstein's GW. The scalar field $\varphi=\varphi_{0}$
has to be added to the Minkowskian background \cite{key-10,key-12}.
We also assume $\varphi_{0}$ to be a minimum for $W$: 

\begin{equation}
W\simeq\frac{1}{2}\alpha\delta\varphi^{2}\Rightarrow W'\simeq\alpha\delta\varphi\label{eq: minimo}\end{equation}

The linearization of equations is parallel to the canonical one of
thesecond Section of this paper, thus one can put

\begin{equation}
\begin{array}{c}
g_{\mu\nu}=\eta_{\mu\nu}+h_{\mu\nu}\\
\\\varphi=\varphi_{0}+\delta\varphi.\end{array}\label{eq: linearizza2}\end{equation}
To first order in $h_{\mu\nu}$ and $\delta\varphi$, the linearized
field equations now are \cite{key-9,key-12}

\begin{equation}
\begin{array}{c}
\widetilde{R}_{\mu\nu}-\frac{\widetilde{R}}{2}\eta_{\mu\nu}=\partial_{\mu}\partial_{\nu}\Phi+\eta_{\mu\nu}[]\Phi\\
\\{}[]\Phi=m^{2}\Phi,\end{array}\label{eq: linearizzate2}\end{equation}

where we have defined

\begin{equation}
\begin{array}{c}
\Phi\equiv-\frac{\delta\varphi}{\varphi_{0}}\\
\\m^{2}\equiv\frac{\alpha\varphi_{0}}{2\omega+3}.\end{array}\label{eq: definizione}\end{equation}

In analogy to the purely General Relativity case of Section 2, the
linearized Riemann tensor and the linearized eqs. (\ref{eq: linearizzate2})
are invariant for gauge transformations

\begin{equation}
\begin{array}{c}
h_{\mu\nu}\rightarrow h'_{\mu\nu}=h_{\mu\nu}-\partial_{(\mu}\epsilon_{\nu)}\\
\\\delta\varphi\rightarrow\delta\varphi'=\delta\varphi;\end{array}\label{eq: gauge massiva}\end{equation}

then one can define

\begin{equation}
\bar{h}_{\mu\nu}\equiv h_{\mu\nu}-\frac{h}{2}\eta_{\mu\nu}+\eta_{\mu\nu}\Phi,\label{eq: ridefiniz massiva}\end{equation}

and, considering the gauge transform (Lorenz condition) with the condition 

\begin{equation}
[]\epsilon_{\nu}=\partial^{\mu}\bar{h}_{\mu\nu}\label{eq:lorentziana massiva}\end{equation}

for the parameter $\epsilon^{\mu}$, it is

\begin{equation}
\partial^{\mu}\bar{h}_{\mu\nu}=0.\label{eq: cond lorentz massiva}\end{equation}

Thus, the field equations like can be rewritten

\begin{equation}
[]\bar{h}_{\mu\nu}=0\label{eq: onda T massiva}\end{equation}

\begin{equation}
[]\Phi=m^{2}\Phi.\label{eq: onda S massiva}\end{equation}

Solutions of eqs. (\ref{eq: onda T massiva}) and (\ref{eq: onda S massiva})
are plan waves:

\begin{equation}
\bar{h}_{\mu\nu}=A_{\mu\nu}(\overrightarrow{p})\exp(ip^{\alpha}x_{\alpha})+c.c.\label{eq: sol T massiva}\end{equation}

\begin{equation}
\Phi=a(\overrightarrow{p})\exp(iq^{\alpha}x_{\alpha})+c.c.\label{eq: sol S massiva}\end{equation}
where

\begin{equation}
\begin{array}{ccc}
k^{\alpha}\equiv(\omega,\overrightarrow{p}) &  & \omega=p\equiv|\overrightarrow{p}|\\
\\q^{\alpha}\equiv(\omega_{m},\overrightarrow{p}) &  & \omega_{m}=\sqrt{m^{2}+p^{2}}.\end{array}\label{eq: k e q}\end{equation}

In eqs. (\ref{eq: onda T massiva}) and (\ref{eq: sol T massiva})
the equation and the solution for the tensorial waves exactly like
in General Relativity have been obtained (Section 2), while eqs. (\ref{eq: onda S massiva})
and (\ref{eq: sol S massiva}) are respectively the equation and the
solution for the massive scalar mode.

Note: now the dispersion law for the modes of the massive scalar field
$\Phi$ is not linear \cite{key-12,key-14,key-15}. The velocity of
every tensorial mode $\bar{h}_{\mu\nu}$ is the light speed $c$,
but the dispersion law (the second of eq. (\ref{eq: k e q})) for
the modes of $\Phi$ is that of a massive field which can be discussed
like a wave-packet \cite{key-12,key-14,key-15}. Also, the group-velocity
of a wave-packet of $\Phi$ centered in $\overrightarrow{p}$ is 

\begin{equation}
\overrightarrow{v_{G}}=\frac{\overrightarrow{p}}{\omega},\label{eq: velocita' di gruppo}\end{equation}

which is exactly the velocity of a massive particle with mass $m$
and momentum $\overrightarrow{p}$.

From the second of eqs. (\ref{eq: k e q}) and eq. (\ref{eq: velocita' di gruppo})
it is simple to obtain \cite{key-12,key-14}:

\begin{equation}
v_{G}=\frac{\sqrt{\omega^{2}-m^{2}}}{\omega}.\label{eq: velocita' di gruppo 2}\end{equation}

Then, wanting a constant speed of our wave-packet, one needs

\begin{equation}
m=\sqrt{(1-v_{G}^{2})}\omega.\label{eq: relazione massa-frequenza}\end{equation}

The relation (\ref{eq: relazione massa-frequenza}) is shown in fig.
3 for a value $v_{G}=0.9$.

\begin{figure}
\includegraphics{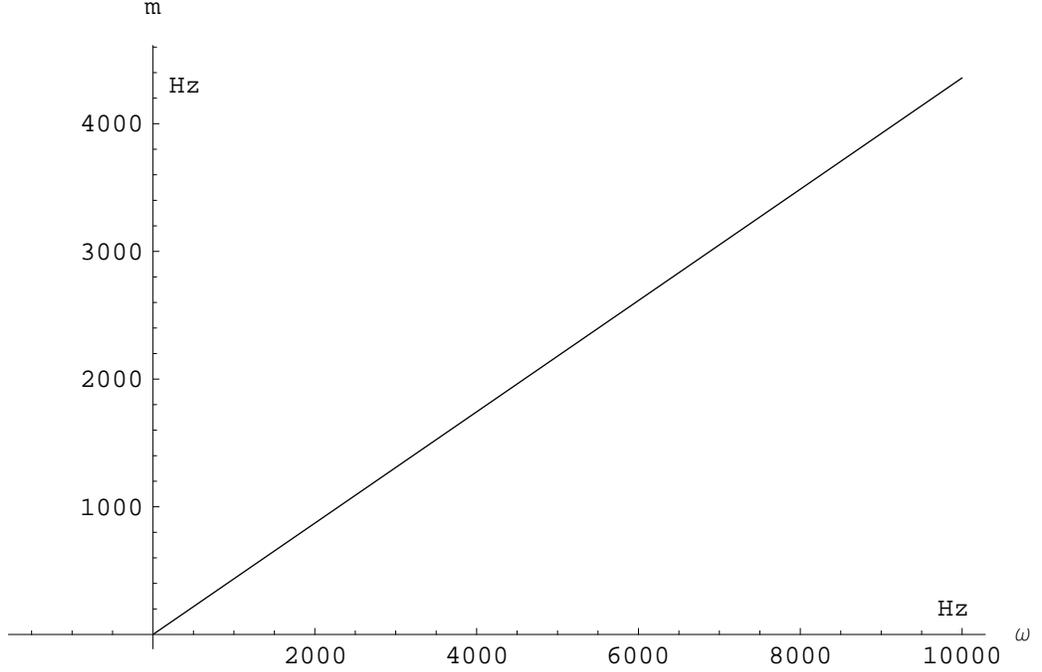}

\caption{from Capozziello S and Corda C - Int. J. Mod. Phys. D \textbf{15}
1119 -1150 (2006), the mass-frequency relation for a massive SGW incoming
with a speed of $0.9c$ : for the mass it is $1Hz=10^{-15}eV$}
\end{figure}

Now let us remain in the Lorenz gauge with trasformations of the type
$[]\epsilon_{\nu}=0$; this gauge gives a transverse condition for
the tensorial part of the field: $k^{\mu}A_{\mu\nu}=0$, but we do
not know if the total field $h_{\mu\nu}$ is transverse \cite{key-15}.
From eq. (\ref{eq: ridefiniz massiva}) it is:

\begin{equation}
h_{\mu\nu}=\bar{h}_{\mu\nu}-\frac{\bar{h}}{2}\eta_{\mu\nu}+\eta_{\mu\nu}\Phi.\label{eq: ridefiniz 2}\end{equation}

At this point, being in the massless case one could put

\begin{equation}
\begin{array}{c}
[]\epsilon^{\mu}=0\\
\\\partial_{\mu}\epsilon^{\mu}=-\frac{\bar{h}}{2}+\Phi,\end{array}\label{eq: gauge2 massiva}\end{equation}

which gives the transverse condition of the total field. But in the
massive case this is impossible. In fact, applying the D'Alembertian
operator to the second of eqs. (\ref{eq: gauge2 massiva}) and using
the field equations (\ref{eq: onda T massiva}) and (\ref{eq: onda S massiva})
it is

\begin{equation}
[]\epsilon^{\mu}=m^{2}\Phi,\label{eq: contrasto}\end{equation}

which is in contrast with the first of eqs. (\ref{eq: gauge2 massiva}).
In the same way it is possible to show that it does not exist any
linear relation between the tensorial field $\bar{h}_{\mu\nu}$ and
the scalar field $\Phi$. Thus a gauge in wich $h_{\mu\nu}$ is purely
spatial cannot be chosen (i.e. we cannot put $h_{\mu0}=0,$ see eq.
(\ref{eq: ridefiniz 2})). But one can put the traceless condition
to the field $\bar{h}_{\mu\nu}$:

\begin{equation}
\begin{array}{c}
[]\epsilon^{\mu}=0\\
\\\partial_{\mu}\epsilon^{\mu}=-\frac{\bar{h}}{2},\end{array}\label{eq: gauge traceless massiva}\end{equation}

which imply

\begin{equation}
\partial^{\mu}\bar{h}_{\mu\nu}=0.\label{eq: vincolo massivo}\end{equation}

Wanting to save the conditions $\partial_{\mu}\bar{h}^{\mu\nu}$ and
$\bar{h}=0,$ one can use transformations like

\begin{equation}
\begin{array}{c}
[]\epsilon^{\mu}=0\\
\\\partial_{\mu}\epsilon^{\mu}=0,\end{array}\label{eq: gauge 3 massiva}\end{equation}

and, taking $\overrightarrow{p}$ in the $z$ direction, it is possible
to choose a gauge in which only $A_{11}$, $A_{22}$, and $A_{12}=A_{21}$
are different to zero. The condition $\bar{h}=0$ gives $A_{11}=-A_{22}$.
Now, putting these equations in eq. (\ref{eq: ridefiniz 2}) it is

\begin{equation}
h_{\mu\nu}(t,z)=A^{+}(t-z)e_{\mu\nu}^{(+)}+A^{\times}(t-z)e_{\mu\nu}^{(\times)}+\Phi(t-v_{G}z)\eta_{\mu\nu}.\label{eq: perturbazione totale massiva}\end{equation}

The term $A^{+}(t-z)e_{\mu\nu}^{(+)}+A^{\times}(t-z)e_{\mu\nu}^{(\times)}$
describes the two standard (i.e. tensorial) polarizations of gravitational
waves which arises from General Relativity, exactly like in Section
2, while the term $\Phi(t-v_{G}z)\eta_{\mu\nu}$ is the massive scalar
field. 

Now it will be shown that the discussion of the physical effect of
the wave is different from the massless case \cite{key-12,key-13,key-14}.

For a pure scalar gravitational wave equation (\ref{eq: perturbazione totale massiva})
becomes \begin{equation}
h_{\mu\nu}(t,z)=\Phi(t-v_{G}z)\eta_{\mu\nu},\label{eq: perturbazione scalare massiva}\end{equation}

and, with an analysis parallel to the one of previous Section, using
equation (\ref{eq: perturbazione totale massiva}) directly in equations
(\ref{eq: riemann lineare}), for the gauge invariance of the linearized
Riemann tensor, one gets in the frame of the local observer \begin{equation}
\begin{array}{c}
\widetilde{R}_{010}^{1}=-\frac{1}{2}\ddot{\Phi}\\
\\\widetilde{R}_{010}^{2}=-\frac{1}{2}\ddot{\Phi}\\
\\\widetilde{R}_{030}^{3}=\frac{1}{2}[]\Phi.\end{array}\label{eq: componenti riemann 2}\end{equation}

But, putting the field equation (\ref{eq: onda S massiva}) in the
third of eqs. (\ref{eq: componenti riemann 2}) it is

\begin{equation}
\widetilde{R}_{030}^{3}=\frac{1}{2}m^{2}\Phi,\label{eq: terza riemann 2}\end{equation}

which shows that the field is not transversal. 

Infact, using eq. (\ref{eq: deviazione geodetiche}) it results

\begin{equation}
\ddot{x}=\frac{1}{2}\ddot{\Phi}x,\label{eq: accelerazione mareale lungo x 2}\end{equation}

\begin{equation}
\ddot{y}=\frac{1}{2}\ddot{\Phi}y\label{eq: accelerazione mareale lungo y 2}\end{equation}

and 

\begin{equation}
\ddot{z}=-\frac{1}{2}m^{2}\Phi(t,z)z.\label{eq: accelerazione mareale lungo z 2}\end{equation}

Then the effect of the mass is the generation of a \textit{longitudinal}
force (in addition to the transverse one). 

For a better understanding of this longitudinal force, let us analyse
the effect on test masses in the context of the geodesic deviation.

Following \cite{key-14} one puts

\begin{equation}
\widetilde{R}_{0j0}^{i}=\frac{1}{2}\left(\begin{array}{ccc}
-\partial_{t}^{2} & 0 & 0\\
0 & -\partial_{t}^{2} & 0\\
0 & 0 & m^{2}\end{array}\right)\Phi(t,z)=-\frac{1}{2}T_{ij}\partial_{t}^{2}\Phi+\frac{1}{2}L_{ij}m^{2}\Phi.\label{eq: eqs}\end{equation}

Here we have used the transverse projector with respect to the direction
of propagation of the GW $\widehat{n}$, defined by

\begin{equation}
T_{ij}=\delta_{ij}-\widehat{n}_{i}\widehat{n}_{j},\label{eq: Tij}\end{equation}

and the longitudinal projector defined by

\begin{equation}
L_{ij}=\widehat{n}_{i}\widehat{n}_{j}.\label{eq: Lij}\end{equation}

In this way the geodesic deviation equation (\ref{eq: deviazione geodetiche})
can be rewritten like

\begin{equation}
\frac{d^{2}}{dt^{2}}x_{i}=\frac{1}{2}\partial_{t}^{2}\Phi T_{ij}x_{j}-\frac{1}{2}m^{2}\Phi L_{ij}x_{j}.\label{eq: TL}\end{equation}

Thus one sees immediately that the effect of the mass present in the
GW generates a longitudinal force proportional to $m^{2}$ which is
in addition to the transverse one. But if $v_{g}\rightarrow1$ in
eq. (\ref{eq: relazione massa-frequenza}) it is $m\rightarrow0$,
and the longitudinal force vanishes. Thus it is clear that the longitudinal
mode arises from the fact that the scalar part of the GW does no propagate
at the speed of light.

In ref. \cite{key-12} the analysis has been generalized to all the
frequencies (eqs. (\ref{eq: eqs}) and (\ref{eq: TL}) are correct
only in the low frequencies approximation) with the computation of
the longitudinal response function of interferometers for massive
scalar GW, which is (eq. (136) of \cite{key-12})

\begin{equation}
\begin{array}{c}
\Upsilon_{l}(\omega)=(1-\frac{1}{v_{P}^{2}})\exp[i\omega(1+\frac{1}{v_{P}})L]+\frac{v_{P}(1-\frac{1}{v_{P}^{2}})}{4L\omega}(\frac{\exp[2i\omega L](iv_{P}^{2}-(v_{P}-1)v_{P}\omega+iL(v_{P}-1)^{2}\omega^{2}}{(v_{P}-1)^{3}}+\\
\\\frac{2\exp[i\omega(1+\frac{1}{v_{P}})L](-2iv_{P}^{2}(3v_{P}^{2}+1)+2(1+L)v_{P}(v_{P}^{4}-1)\omega+iL^{2}(v_{P}+1)^{2}\omega^{2})}{(v_{P}^{2}-1)^{3}}+\\
\\-\frac{2iv_{P}^{2}+2v_{P}(v_{P}+1)\omega+2iL(v_{P}+1)^{2}\omega^{2}}{(v_{P}+1)^{3}}),\end{array}\label{eq: risposta totale lungo z due}\end{equation}

and for $m\rightarrow0$ (i.e. $v_{P}\rightarrow1$) one gets $\Upsilon_{l}(\omega)\rightarrow0$.

In figure 4 the longitudinal response function of the Virgo interferometer
to two massive scalar GWs with speeds of $0.1c$ (non relativistic
case) and $0.9999c$ (ultra relativistic case, thick line) is shown.
In the non-relativistic case the longitudinal signal is strong while
in the ultra relativistic case (i.e. $v_{P}\rightarrow1$) is very
weak. %
\begin{figure}
\includegraphics{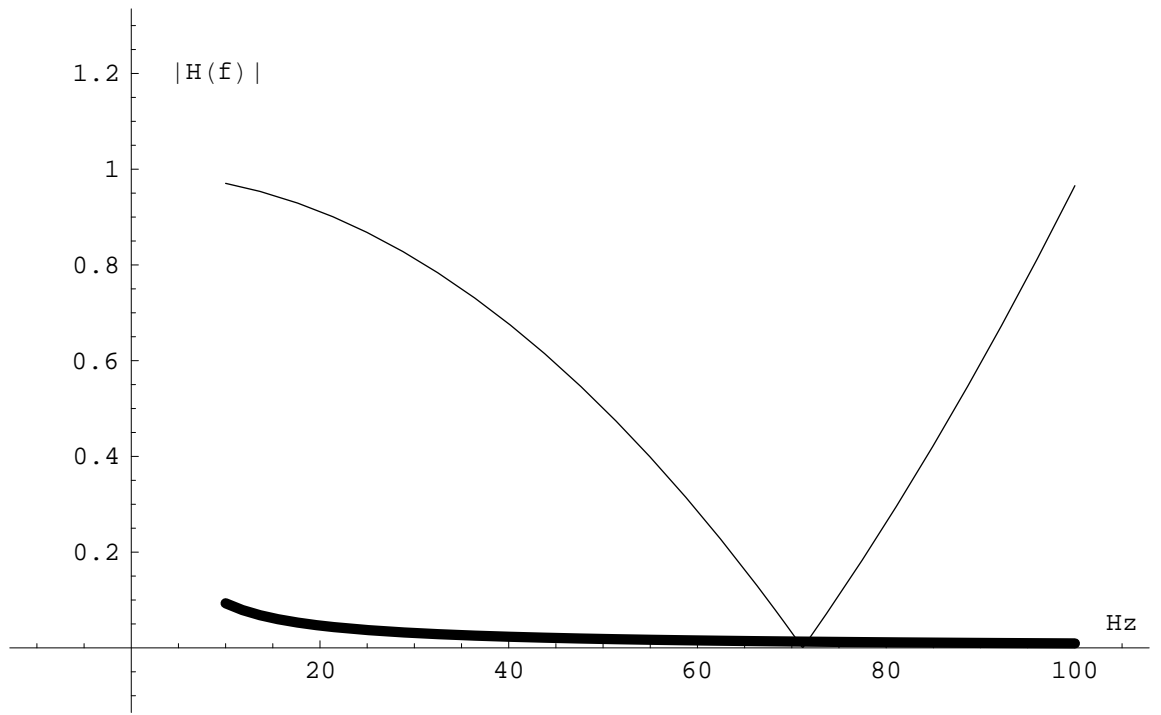}

\caption{from Capozziello S and Corda C - Int. J. Mod. Phys. D \textbf{15}
1119 -1150 (2006), the absolute value, at low frequencies, of the
longitudinal response function of the Virgo interferometer to two
SGWs with speeds of $0.1c$ (non relativistic case) and $0.9999c$
(ultra relativistic case, thick line).}
\end{figure}

\section{A control of gauge invariance: computation in the TT gauge for Einstein's
GWs}

In this Section, returning in the context of standard Einstein's GWs,
a brief review of the portion of ref. \cite{key-20}, which analyzes
the TT gauge, is given for a control of gauge invariance that shows
the correctness of the results.

Let us start by considering the interval for a photon which propagates
in the $x$ axis. With the same considerations about the conservation
of the photon momentum, one can write eq. (\ref{eq: metrica +}) as

\begin{equation}
ds^{2}=-dt^{2}+[1+h(t-z)]dx^{2}.\label{eq: metrica l}\end{equation}
Putting the condition $ds^{2}=0$ for null geodesics in the line element
(\ref{eq: metrica l}) the coordinate velocity of the photon is obtained

\begin{equation}
v^{2}\equiv(\frac{dx}{dt})^{2}=\frac{1}{1+h(t)},\label{eq: velocita' fotone}\end{equation}

which is a convenient quantity for calculations of the photon propagation
time between the beam-splitter and the mirror \cite{key-2,key-12,key-15,key-20}.
In the TT gauge, the time coordinate $t$ is also the proper time
at every point of the space, because $g_{00}=1$ \cite{key-21,key-23}.

In this gauge the coordinates of the beam-splitter $x_{b}=l$ and
of the mirror $x_{m}=l+L_{0}$ do not changes under the influence
of the GW \cite{key-2,key-20}, thus one can find the duration of
the forward trip as

\begin{equation}
T_{1}(t)=\int_{l}^{L_{0}+l}\frac{dx}{v(t')},\label{eq: durata volo}\end{equation}

with 

\begin{center}$t'=t-(l+L_{0}-x)$,\end{center}

where in the last equation $t'$ is the retardation time. 

To first order in $h$ this integral can be approximated with

\begin{equation}
T_{1}(t)=T_{0}+\frac{1}{2}\int_{l}^{L_{0}+l}h(t')dx,\label{eq: durata volo andata approssimata}\end{equation}

where

\begin{center}$T_{0}=L_{0}$ \end{center}

is the transit time of the photon in absence of the GW. Similiary,
the duration of the return trip will be\begin{equation}
T_{2}(t)=T_{0}+\frac{1}{2}\int_{l+L_{0}}^{l}h(t')(-dx),\label{eq: durata volo ritorno approssimata}\end{equation}

though now the retardation time is 

\begin{center}$t'=t-(x-l)$.\end{center}

Equations (\ref{eq: durata volo andata approssimata}) and (\ref{eq: durata volo ritorno approssimata})
are exactly equal to equations (\ref{eq: durata volo andata approssimata in Corda x})
and (\ref{eq: durata volo ritorno approssimata in Corda x}). Then,
the same computation that has been made in Section 4 can be performed
in this case too, obtaining

\begin{equation}
\delta T(t)=\frac{1}{2}\int_{l}^{L_{0}+l}[h(t-2L_{0}+x-l)+h(t-x+l)]dx.\label{eq: variazione temporale in gauge Corda  lungo x}\end{equation}

Thus, in the $x$ direction of the TT gauge, there is the same variation
of proper time (distance) of eq. (\ref{eq: variazione temporale in gauge comovente}),
which was obtained in the $x$ direction of the different gauge analyzed
in Section 4. 

For a photon which propagates in the $z$ axis the analysis is trivial.
The condition $ds^{2}=0$ for null geodesic gives now 

\begin{equation}
dt=\pm dz.\label{eq: metrica TT lungo z 2}\end{equation}

Then, by recalling that $t$ is the proper time in this gauge, it
is:

\begin{equation}
\tau(t)=T_{0}+T_{0}=2T_{0}.\label{eq: tempo proprio totale lungo z in Corda2}\end{equation}

Thus it is $\delta\tau=\delta L_{0}=0$ , i.e. a longitudinal effect
is not present. 

By confronting eq. (\ref{eq: variazione temporale in gauge Corda  lungo x})
with eq. (\ref{eq: variazione temporale in gauge comovente}) and
eq. (\ref{eq: tempo proprio totale lungo z in Corda2}) with eq. (\ref{eq: tempo proprio totale lungo z in Corda})
the gauge invariance between our gauge and the TT one is recovered.

\section{Conclusions }

Because the investigation of the transverse effect of gravitational
waves (GWs) could constitute a further tool to discriminate among
several relativistic theories of gravity on the ground, after a review
of the TT gauge, the transverse effect of Einstein's GWs has been
reanalized with a different choice of coordinates. In the chosen gauge
test masses have an apparent motion in the direction of propagation
of the wave, while in the transverse direction they appear at rest.
Of course, this is only a gauge artefact. In fact, from careful investigation
of this particular gauge, it has been shown that the tidal forces
associated with GWs act along the directions orthogonal to the direction
of propagation of waves. In the analysis it has also been shown, in
a heuristic way, that the tranverse effect of Einstein's GWs arises
from the propagation of the waves at the speed of light, thus only
massless GWs (and this is the case of Einstein's ones) are purely
transverse (i.e. the presence of the mass precludes GWs to propagate
at the speed of the light). But, because the physics of gravitational
waves has to be investigated by studing the tidal forces as appearing
in the geodesic deviation equation and directly in a laboratory enviroment
on Earth, an analysis of these tidal forces and of the transverse
effect in the frame of the local observer has also been performed.
After this, for a further better understanding of the transverse effect,
an example of a wave, which arises from scalar tensor gravity, with
both transverse and genuinely longitudinal modes has been given and
discussed. In the example the connection between the longitudinal
component and the velocity of the wave has been mathematical shown.

At the end of this paper the review of the TT gauge has been completed,
recovering the gauge invariance between the presented gauge and the
TT one.

\section*{Acknowledgements}

I would like to thank Salvatore Capozziello, Mauro Francaviglia, Maria
Felicia De Laurentis and Giancarlo Cella for helpful advices during
my work. It is a pleasure to thank Giampiero Esposito for clarifications
about the Lorenz gauge. I thank the referee for precious advices and
comments that allowed to improve this paper. The European Gravitational
Observatory (EGO) consortium has also to be thanked for the using
of computing facilities.

\end{document}